\documentclass[pra,aps,twocolumn,floatfix,showpacs,tightenlines,superscriptaddress]{revtex4}
\usepackage{amssymb,amsmath,amsthm,amsbsy,epsfig,color,graphicx,times,bbm}
\usepackage[ansinew]{inputenc}
\usepackage[english]{babel}
\usepackage{accents}
\usepackage{color}
\usepackage{braket}
\usepackage{amsfonts}
\usepackage{booktabs}
\usepackage{tabularx}
\usepackage{array}
\usepackage{hyperref}
\hypersetup{pdfstartview=FitH}
\usepackage{latexsym}
\usepackage{amsthm}
\newtheorem{theorem}{Theorem}
\newcommand{\virg}[1]{``#1''}

\begin{document}

\title{n-cluster models in a transverse magnetic field}

\author{G. Zonzo}
\thanks{Corresponding author: giuseppezonzo@gmail.com}
\affiliation{Dipartimento di Fisica, Universit\`a degli Studi di Salerno, Via Giovanni Paolo II 132, I-84084 Fisciano (SA), Italy}

\author{S. M. Giampaolo}
\affiliation{International Institute of Physics, UFRN, Av. Odilon Gomes de Lima 1722, 59078-400 Natal, Brazil}

%

\begin{abstract}
In this paper we analyze a family of one dimensional fully analytically solvable models, named the $n$-cluster models in a transverse magnetic field, 
in which a many-body cluster interaction competes with a uniform transverse magnetic field. 
These models, independently by the cluster size $n\!+\!2$, exhibit a quantum phase transition, that separates a paramagnetic phase from a cluster one, 
that corresponds to a nematic ordered phase or a symmetry-protected topological ordered phase for even or odd $n$ respectively. 
Due to the symmetries of the spin correlation functions, we prove that these models have no genuine multipartite entanglement. 
On the contrary, for a magnetic field strong enough, a non vanishing concurrence arises between spins at the endpoints of the cluster. 
Due to their integrability and entanglement properties, the $n$-cluster models in a transverse magnetic field may serve as a prototype for studying 
non trivial-spin orderings and as a potential reference system for the applications of quantum information tasks.
\end{abstract}


\maketitle

\section{Introduction}

Accordingly with the Ginzburg-Landau approach, the appearance of an ordered phase in a classical many-body system is associated to the rising of a 
local order parameter that breaks some symmetries of the Hamiltonian. 
However, this picture is unsuitable to describe all possible kind of orders in a complex quantum system. 
A paradigmatic example are the translation invariant spin-$1/2$ chains, for which the ground states correspond to the so called valence bond states, 
i.e. states made by tensor products of Bell states~\cite{A1973,A1987}. 
In such a case, any possible local operator, i.e. any operator with a support on a single spin of the system, shows a zero expectation value and hence 
there is no local order parameter. 
Nevertheless it is impossible to negate that some kind of order is present in the system.

Among the different kind of ordered phases without a classical counterpart, the nematic and the topological ones are attracting an increasing interest. 
Nematic phases~\cite{AG1984, LMM2011} occur if it is possible to define a ground state that: 
a) It breaks at least one symmetry of the Hamiltonian; 
b) It is characterized by an order parameter with a support on a finite set of sites with dimension is strictly greater than one single spin.
In this sense, the nematic order can be seen as a generalization of the magnetic one to the case in which the order parameter is not strictly local. 
On the opposite limit, the topological ordered phases are characterized by string order parameters, i.e. non vanishing expectation value of operators 
which support extends on the whole system.~\cite{MDSS1996,DQ2012}.   

These novel phases are extremely relevant both from theoretical and applicative point of view. 
In fact, topological ordered phases are associated to the robustness of ground state degeneracies~\cite{WN1990}, show quantized non-Abelian geometric 
phases~\cite{W1990}, possess peculiar patterns of long-range quantum entanglement~\cite{CGW2010}, play a fundamental role in the spin 
liquids~\cite{IHM2011,ZGV2011} and in non-Abelian fractional Hall systems~\cite{LH2008}.
Moreover they are predicted to play a key role in the future development of fault-tolerant quantum computers~\cite{K2003}. 
On the other hand, the nematic order is usually found in materials commercially used in the liquid crystal technology~\cite{TGC2010}.

Due to the great interest on these novel phases the works on many body systems in which they are present is continuously increasing.
To limit ourselves to the one dimensional spin systems, it is known that the frustrated one dimensional spin-$1/2$  chain in an external magnetic field 
shows a nematic ordered phase~\cite{C1991,HHV2006}, the one dimensional cluster Ising model exhibits a symmetry-protected topological ordered 
phase~\cite{SAFFFPV2011,MH2012,GH2014} and the $n$-cluster Ising models, that can obtained using Floquet interactions in atomic systems~\cite{LJR2016},
show both nematic and topological orders, depending on $n$~\cite{GH2015}.
Following this line of research, the goal of this paper is to provide a complete analysis of a set of one dimensional fully analytically spin-$1/2$
solvable models, that falls into different classes of symmetry. The Hamiltonian of these models can be written as
 \begin{equation}
  \label{Hamiltonian}
 H_{\phi}^{(n)}\!=\! -J \cos(\phi)\!\sum_{j}\! \sigma_{j}^x O^z_{j,n} \sigma_{j+n+1}^x\!+\!J \sin(\phi)\!  \sum_{j}\! \sigma_{j}^z \; ,
  \end{equation}
where $\phi$ controls the relative weight of the interactions, $J$ has the dimension of an energy, $\sigma_i^\alpha$ (with  $\alpha=x,y,z$) are the 
Pauli operators and $O^z_{j,n}$ stands for
\begin{equation}
\label{Oz_operator}
 O^z_{j,n} = \bigotimes_{k=1}^{n} \sigma_{j+k}^z \; .
\end{equation}
We show that such models can be diagonalized (Sec.~\ref{Solution}) and all the spin correlation functions can be analytically obtained 
(Sec.~\ref{Spincorrelation}) using the Jordan-Wigner transformations~\cite{JW1928}. 
We prove that, independently by the length of $O^z_{j,n}$, all the models show a quantum critical point at $\phi_c=\pi/4$, when both the local and 
the cluster interactions have the same weight. 
For $\phi>\phi_c$ the system is in a paramagnetic phase, while for $\phi<\phi_c$, the system exhibits a cluster phase, that is a nematic or 
topological ordered phase for even or odd $n$ respectively. 
In both cases we determine the order parameter, i.e. a string order parameter for the topological ordered phase and a block order parameter for the 
nematic one (Sec.~\ref{Orderparameters}).

In Sec.~\ref{Entanglementproperties}, we analyze the entanglement properties of such models. 
At first, we focus on the entanglement between two spins in a block, as quantified by the concurrence. 
In contrast with the $n$-cluster Ising model~\cite{GH2015}, for any value of $n$, there is a region of the parameter $\phi$ for which the 
entanglement between a pair of spins does not vanish. 
On the other hand, the multipartite entanglement that characterize the $n$-cluster Ising model is completely absent from the model under analysis so
proving that, the presence of a cluster interactions is not sufficient to generate multipartite entanglement in a system.
Then, we analyze the behavior of the block entanglement, at the quantum phase transition $\phi=\phi_c\equiv\pi/4$ and by using the conformal field 
theory~\cite{CC2004} we evaluate the central charge of the models, that turns out to be dependent on $n$, proving that the models fall into 
different classes of symmetry.
In Sec.~\ref{Conclusions}, we draw our conclusions.

\section{Solution of the models}
\label{Solution}

In this section, we show how it is  possible to diagonalize the systems described in eq. (\ref{Hamiltonian}), by following the well known approach 
based on the Jordan-Wigner transformation~\cite{JW1928}. 
The idea at the basis of the Jordan-Wigner approach is to map the Hamiltonian of one dimensional spin-$1/2$ particles into a non-interacting 
spinless fermions moving in the chain~\cite{LSM1961,BM1971,BMD1970,SAFFFPV2011,GH2015}.

Taking into account the local raising and lowering operators $\sigma_j^\pm=(\sigma_j^x \pm \imath \sigma_j^y)/2$, one associates non-local 
fermionic operators to local spin operators
\begin{eqnarray}
\label{JWT}
 c_j = \bigotimes_{k=1}^{j-1}\left( \sigma^z_k \right) \sigma_j^- \;, \;\; \;& & \; \; \;
 c_j^\dagger = \bigotimes_{k=1}^{j-1} \left( \sigma^z_k \right) \sigma_j^+ \; ,
\end{eqnarray}
so that the Hamiltonian in eq.~(\ref{Hamiltonian}) takes the form
\begin{eqnarray}\label{Hamiltonian2}
  H_{\phi}^{(n)}\!\!&\!=\!&\!J\!\cos(\phi) \!\sum_{j}\!\left( c_{j}^\dagger c_{j+n+1}
  + c_{j}^\dagger c_{j+n+1}^\dagger + h.c. \right) \nonumber \\
  &\!-\!&\! J\! \sin(\phi)  \! \sum_{j}\! \left( \! 2 c_{j}^\dagger c_{j} - 1 \right) 
  \end{eqnarray}
We note that the non locality of the Jordan-Wigner transformations maps the cluster interaction, that involves an interaction among $n+2$ spins, 
into a fermionic term between sites at a distance $n + 1$. 

Before to proceed we have to underline a relevant point. 
Looking at the Hamiltonian in eq.~(\ref{Hamiltonian2}) one can be tempted to re-arrange the different terms in such a way that the model under 
analysis can be seen as a set of independent fermionic problems on one dimensional ring with hopping between the nearest sites. 
But this re-arrangement does not take into account the Jordan-Wigner transformations in eq.~(\ref{JWT}) that strongly depends on the order of the 
operators. 
Hence, coming back to the spin models after the re-arrangement, would bring the system in a model that is completely different from the starting one.

To diagonalize the fermionic Hamiltonian in eq.~(\ref{Hamiltonian2}) we perform a Fourier transform
\begin{eqnarray}
\label{FT1}
 b_k & = & \frac{1}{\sqrt{N}} \sum_{j} c_k\; e^{-i\,kj}\; , \nonumber \\
 b_k^\dagger & = & \frac{1}{\sqrt{N}} \sum_{j} c_k^\dagger\; e^{i\,kj} \; ,
\end{eqnarray}
where the wave number $k$ is equal to $k=2 \pi l/N$ and $l$ runs from $-N/2$ to $N/2$, being $N$ the total number of spins (sites) in the chain. 
Thanks to the Fourier transform, the Hamiltonian of eq.~(\ref{Hamiltonian2}) can be written as the sum of $N/2$ non interacting terms
\begin{eqnarray}
  \label{Hamiltonian3}
 H_{\phi}^{(n)} & = & \sum_{k>0} \tilde{H}_{\phi,k}^{(n)} \; .
 \end{eqnarray}
where $\tilde{H}_{\phi,k}^{(n)}$ acts only on fermions with momentum $k$ and $-k$. 
This local Hamiltonian reads 
 \begin{eqnarray}
 \tilde{H}_{\phi,k}^{(n)}& =& 2\,i\, \delta_{\phi,k}^{(n)}\;  \left( b_{k}^\dagger b_{-k}^\dagger- b_{-k} b_{k}\right) \nonumber\\
 &&+
 2\, \varepsilon_{\phi,k}^{(n)}\; \left( b_{k}^\dagger b_{k}+ b_{-k}^\dagger b_{-k}-1 \right) \; ,
\end{eqnarray}
with the parameters $\delta_{\phi,k}^{(n)}$ and $\varepsilon_{\phi,k}^{(n)}$ given by
\begin{eqnarray}
  \label{deltaepsilon}
 \delta_{\phi,k}^{(n)} & = & J\;\sin\left((n+1)k\right) \cos\phi\;, \nonumber \\
 \varepsilon_{\phi,k}^{(n)} & = &  J\;\left(\cos\left((n+1)k\right) \cos\phi + \sin\phi\right)\; .
\end{eqnarray}
In the occupation number basis $|1_k,1_{-k}\rangle$, $|0_k,0_{-k}\rangle$, $|1_k,0_{-k}\rangle$, 
$|0_k,1_{-k}\rangle$, each $\tilde{H}_{\phi,k}^{(n)}$ corresponds to a four level system represented by the following matrix
\begin{equation}
 \label{Hamiltoniank}
\tilde{H}_{\phi,k}^{(n)}=\left(
\begin{array}{cccc}
2\; \varepsilon_{\phi,k}^{(n)} & 2\, i\; \delta_{\phi,k}^{(n)} & 0 & 0 \\
- 2\, i\; \delta_{\phi,k}^{(n)} & - 2\; \varepsilon_{\phi,k}^{(n)} & 0 & 0 \\
 0 & 0 & 0 & 0 \\
 0 & 0 & 0 & 0
\end{array}
\right) \; .
\end{equation}
The ground state energy can be easily computed to be equal to
\begin{eqnarray}
\label{Ek0}
 E_{\phi,k}^{(n)}&=&- 2 \sqrt{\left(\varepsilon_{\phi,k}^{(n)}\right)^2+\left(\delta_{\phi,k}^{(n)}\right)^2} \nonumber \\
 &=&- 2J \sqrt{1+\cos((n+1)k) \sin(2\phi)} \;,
\end{eqnarray}
while the associated ground state $|\psi_{\phi,k}^{(n)}\rangle$ is represented by the superposition of $|1_k,1_{-k}\rangle$ and $|0_k,0_{-k}\rangle$
\begin{equation}
|\psi_{\phi,k}^{(n)}\rangle=\alpha_{\phi,k}^{(n)}\; |1_k,1_{-k}\rangle +\beta_{\phi,k}^{(n)}\; |0_k,0_{-k}\rangle \; ,
\end{equation}
with
\begin{eqnarray}
 \label{alphakbetak}
 \alpha_{\phi,k}^{(n)} &=&i \frac{\varepsilon_{\phi,k}^{(n)} -E_{\phi,k}^{(n)}}{\sqrt{\left(\delta_{\phi,k}^{(n)}\right)^2
 +\left(\varepsilon_{\phi,k}^{(n)}
 -E_{\phi,k}^{(n)}\right)^2}} \; ,
 \nonumber \\
 \beta_{\phi,k}^{(n)} &=&  \frac{\delta_{\phi,k}^{(n)}}{\sqrt{\left(\delta_{\phi,k}^{(n)}\right)^2+\left(\varepsilon_{\phi,k}^{(n)}
 -E_{\phi,k}^{(n)}\right)^2}} \; .
\end{eqnarray}

Since the Hamiltonian is the sum of the non-interacting terms $\tilde{H}_{\phi,k}^{(n)}$, each one of them acting onto a different Hilbert space,
the ground state of the total Hamiltonian is the tensor product of all $|\psi_{\phi,k}^{(n)}\rangle$
\begin{equation}
 |\psi_{\phi}^{(n)}\rangle=\bigotimes_{k} |\psi_{\phi,k}^{(n)}\rangle\;,
\end{equation}
and the associated energy density $E_{\phi}^{(n)}$ is the mean of $E_{\phi,k}^{(n)}$ over the total number of spins $N$.

\begin{figure}
\includegraphics[width=0.42\textwidth]{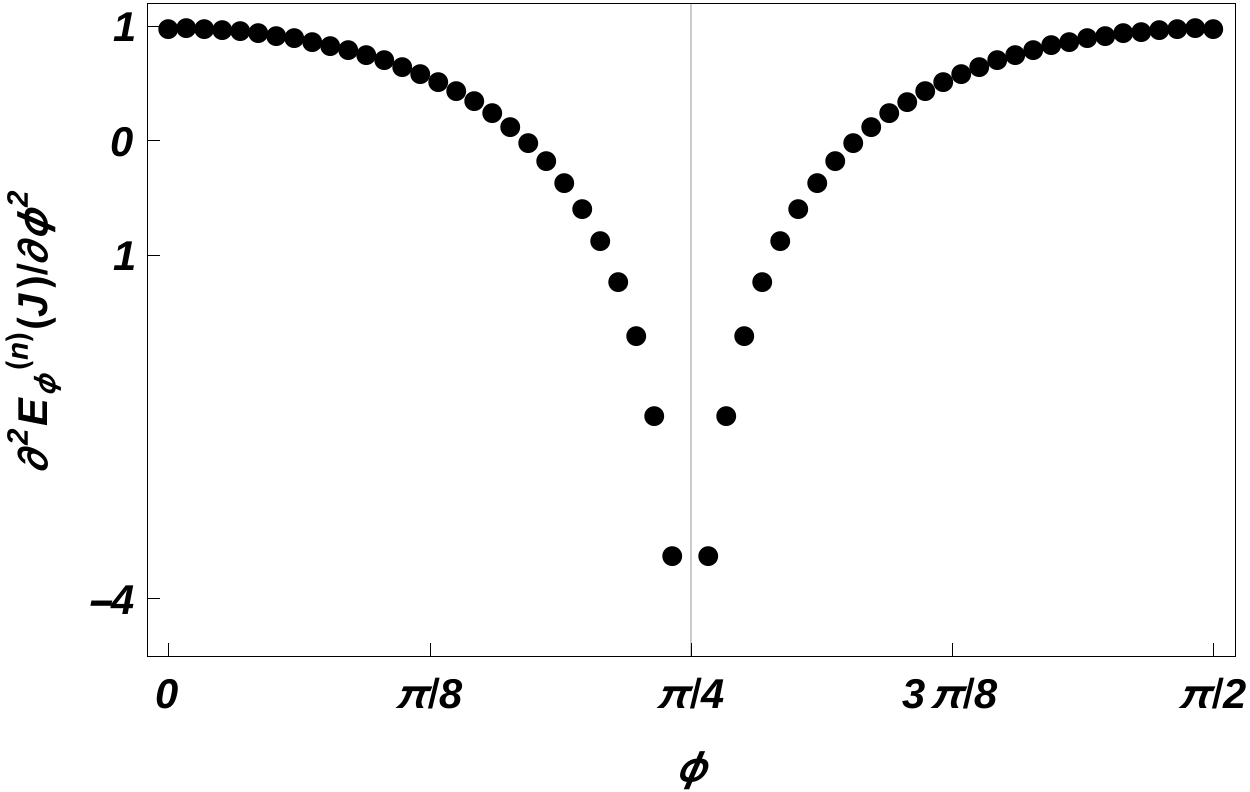}
\caption{(Color online) Behavior of the second derivative of the ground state energy density $E_{\phi,k}^{(n)}$, as function of $\phi$, for 
$n=1$. The divergence is independent of $n$ at the critical value $\phi_c=\frac{\pi}{4}$ and corresponds to a vanishing 
energy gap between the ground state and the first excited state.}
\label{secondderivativeenergy}
\end{figure}

In the thermodynamic limit, replacing the sum with the integral, the expression of the energy density becomes
\begin{equation}
 \label{E0}
E_{\phi}^{(n)}=-\frac{2J}{\pi} \int_0^\pi  \sqrt{1+\cos((n+1)k) \sin(2\phi)} dk \;.
\end{equation}
According to the general theory of continuous phase transitions at zero temperature~\cite{S2011}, the divergence of the second derivative of the 
energy density, with respect to the Hamiltonian parameter, signals the presence of a quantum critical point. In Fig.~(\ref{secondderivativeenergy}), 
we plot the second derivative of the energy density as a function of $\phi$. It clearly shows a divergence at $\phi=\phi_c\equiv\pi/4$, 
independently of $n$.  The singularity corresponds to a vanishing energy gap between the ground and 
the first excited state with the modes $k=\frac{j\pi}{n+2}$, where $j$ runs from $0$ to $n+1$ .

\section{The spin correlations functions}
\label{Spincorrelation}

With the analytic expression of the ground state, any possible spin correlation function can be obtained. 
However, before the evaluation, it is necessary to transform the spin operators into fermionic ones, because the ground state is expressed in 
terms of the fermionic variables. 
It is possible to prove~\cite{BM1971} that all the spin operators can be mapped into ordered products of $2$ types of Majorana fermionic 
operators, indicated with $A_j$ and $B_j$
\begin{equation}
 \label{A&B}
 A_j=c_j+c_j^\dagger \qquad\textrm{and}\quad  B_j=c_j-c_j^\dagger\;,
\end{equation}
where $j$ runs over all the spins of the system. 
Moreover, applying the Wick's theorem~\cite{W1950}, any operator written as product of operators $A_j$ and $B_j$ can be obtained as combination
of one- and two-body expectation values.
Taking into account eq.~(\ref{alphakbetak}), we obtain
\begin{eqnarray}
\label{correlations}
 \langle A_i \rangle & = &  0\;, \nonumber \\
 \langle B_i \rangle & = &  0\;, \nonumber \\
 \langle A_i A_k\rangle  & = & \delta_{ik}\;, \nonumber \\
 \langle B_i B_k\rangle  & = & -\delta_{ik}\;,  \\
 \langle B_i A_k\rangle  & = & G_{i,k}(n,\phi)\;. \nonumber
\end{eqnarray}
where $\braket{\mathcal{O}}$ is a shortcut for $\braket{\psi_{\phi}^{(n)}|\mathcal{O}|\psi_{\phi}^{(n)}}$. 
As a consequence, if a spin operator is mapped into an odd number of fermionic operators, its expectation value on the ground state of the system has
to vanish, because, due to the Wick's theorem it can be written as a combination of terms each one of them including an expectation value of a single 
fermionic operator.
This is the case of spin operators that explicitly break the parity symmetry along $z$. 
Moreover, also spin operators that are mapped in fermionic one with a different number of $A_i$ and $B_i$ operators has to vanish due to the third and 
fourth of eqs.~(\ref{correlations}). 
This is, for example the case of operators like $\sigma^x_i \sigma^y_{j}$.

In all the other case, the spin correlation functions can be expressed in terms of the $G_{i,k}(n,\phi)$ functions. 
Because we are considering models that are invariant under spatial translation, $G_{i,k}(n,\phi)$ depends only on the relative distance 
\mbox{$r=i-k$}, i.e. $G_{i,k}(n,\phi)=G_{r}(n,\phi)$. 
Taking into account eq.~(\ref{alphakbetak}) in the thermodynamic limit, we obtain
\begin{equation}
\label{gfunc}
 \!\!\!\!\!\!G_r(n,\phi)\!=\!\frac{1}{\pi}\!\! \int_0^\pi\!\!
\frac{\cos(k(n\!+\!1\!+\!r))\!\cos\phi\!-\!\cos(kr)\! \sin\phi}{\sqrt{1+\cos((n+1)k)\! \sin(2\phi)}} \!dk 
\end{equation}
To analyze some properties of the $G_r(n,\phi)$ let us define an useful function
\begin{eqnarray}
\label{Clusterfield3-4}
  I(p) \!& \!= \!&\! \frac{1}{\pi} \int_0^\pi \frac{\cos(k p)}{\sqrt{1-\sin(2 \phi)\cos(k(n+1))}} dk \, ; \nonumber \\
  I(p)\! & \!=\! &\! \frac{1}{2 \pi} \int_0^{2\pi} \frac{\cos(k p)}{\sqrt{1-\sin(2 \phi)\cos(k(n+1))}} dk\, .
\end{eqnarray}
In terms of $I'(p)$ the $G_\phi(r)$ can be written as
\begin{equation}
 \label{Clusterfield3-5}
 G_\phi(r)=\cos(\phi)I(n+1+r)+\sin(\phi)I(r)\,.
\end{equation}
We can rewrite the $I(p)$ function as 
\begin{eqnarray}
  \label{Clusterfield3-6}
  \!\!I(p)\! \!&\!=\!&\!\!\frac{1}{2 \pi} \sum_{l=0}^{n}\int_{\frac{2 \pi l}{n+1}}^\frac{2 \pi (l+1)}{n+1}  \frac{\cos(k p)}{\sqrt{1-\sin(2 \phi)\cos(k(n+1))}}
  dk \, ; \nonumber \\
\!\!\!\!\!  &=& \!\!\!\frac{1}{2 \pi} \sum_{l=0}^{n}\int_{0}^\frac{2 \pi}{n+1} \!\!
   \frac{\cos(k p) \cos(\frac{2 \pi l p}{n+1})\!-\!\sin(k p) \sin(\frac{2 \pi l p}{n+1})}{\sqrt{1-\sin(2 \phi)\cos(k(n+1))}} dk\, ;
  \nonumber
 \end{eqnarray}
but the sum $\sum_{l=0}^{n}(\cos(k p) \cos(\frac{2 \pi l p}{n+1})-\sin(k p) \sin(\frac{2 \pi l p}{n+1}))$ is non zero if and only if $p=(n+1)m$
where $m$ is an integer (positive or negative).
Hence $I(p)$ is non zero if and only if $p=(n+1)m$ and consequently also $G_r(n,\phi)$ is non vanishing if and only if 
$r=m(n+1)$ where $m$ in an integer.
This fact plays a fundamental role in the behavior of the entanglement properties among different spins  as we will see in 
Sec.~\ref{Entanglementproperties}.

Knowing $G_r(n,\phi)$ we can make some consideration about the different correlation functions. To begin, let us consider the magnetization along the 
direction of the parity symmetry.
The presence of the external field along the $z$ axis induces a non vanishing magnetization given by 
\begin{equation}
 \langle \sigma_i^z\rangle= -G_{0}(n,\phi) \; ,
\end{equation}
that is always different from zero, for all possible values of $\phi \neq 0$ and for all $n$. 
Moreover, in the same direction, the two-body correlation function can be written as 
\begin{equation}
 \langle \sigma_i^z \sigma_{i+r}^z\rangle= G_{0}^2(n,\phi)-G_{r}(n,\phi)G_{-r}(n,\phi) \;.
\end{equation}
and hence for $r\neq m(n+1)$ it factorizes in the product of the two magnetizations.

Otherwise, if $\mu=x,y$ the spin correlation functions are given by the Slater determinants
\begin{equation}
 \label{correlationmatrixx}
\!\! \!\!\!\!\langle \sigma_i^x \sigma_{i+r}^x\rangle\!=\!
 \left(
 \begin{array}{cccc}
 \! G_{-1}(n,\phi) & G_{-2}(n,\phi) &\! \cdots \!& \!\!G_{-r}(n,\phi)\! \\
 \! G_{0}(n,\phi) & G_{-1}(n,\phi) & \!\cdots \!& G_{1-r}(n,\phi)\!\! \\
 \! \vdots & \vdots & \!\ddots \!& \vdots\! \\
 \! G_{r-2}(n,\phi) &\!\! G_{r-3}(n,\phi) \!\!& \!\cdots\! &\!\! G_{-1}(n,\phi)\!
 \end{array}
\right)\;,
\end{equation}eq.~(\ref{Hamiltonian2})
\begin{equation}
 \label{correlationmatriyy}
\!\! \!\!\!\!\langle \sigma_i^y \sigma_{i+r}^y\rangle\!=\!
 \left(
 \begin{array}{cccc}
 \! G_{1}(n,\phi) & G_{2}(n,\phi) &\! \cdots \!& \!\!G_{r}(n,\phi)\! \\
 \! G_{0}(n,\phi) & G_{1}(n,\phi) & \!\cdots \!& \!\!G_{r-1}(n,\phi)\! \\
 \! \vdots & \vdots & \!\ddots \!& \vdots\! \\
 \! G_{2-r}(n,\phi) & G_{3-r}(n,\phi) & \!\cdots\! & \!\!G_{1}(n,\phi)\!
 \end{array}
\right)\;.
\end{equation}
For all \mbox{$r \neq l(n+1)$}, taking into account that $G_{r}(n,\phi)$ vanishes, we have that 
\begin{eqnarray}
 \label{x&ycorr} 
 \langle \sigma_i^x \sigma_{i+r}^x\rangle &=& 
 \langle \sigma_i^y \sigma_{i+r}^y\rangle = 0 \; .
\end{eqnarray}
In the very relevant case in which $r=n+1$, we obtain that
\begin{eqnarray}
 \langle \sigma_i^x \sigma_{i+n+1}^x\rangle&=&(-1)^n \; G_{-(n+1)}(n,\phi) \; G_0(n,\phi)^n \; , \nonumber \\
 \langle \sigma_i^y \sigma_{i+n+1}^y\rangle&=&(-1)^n \; G_{(n+1)}(n,\phi) \; G_0(n,\phi)^n  \; .
\end{eqnarray}

\section{The order parameters}
\label{Orderparameters}

\begin{figure}[t]
\includegraphics[width=0.42\textwidth]{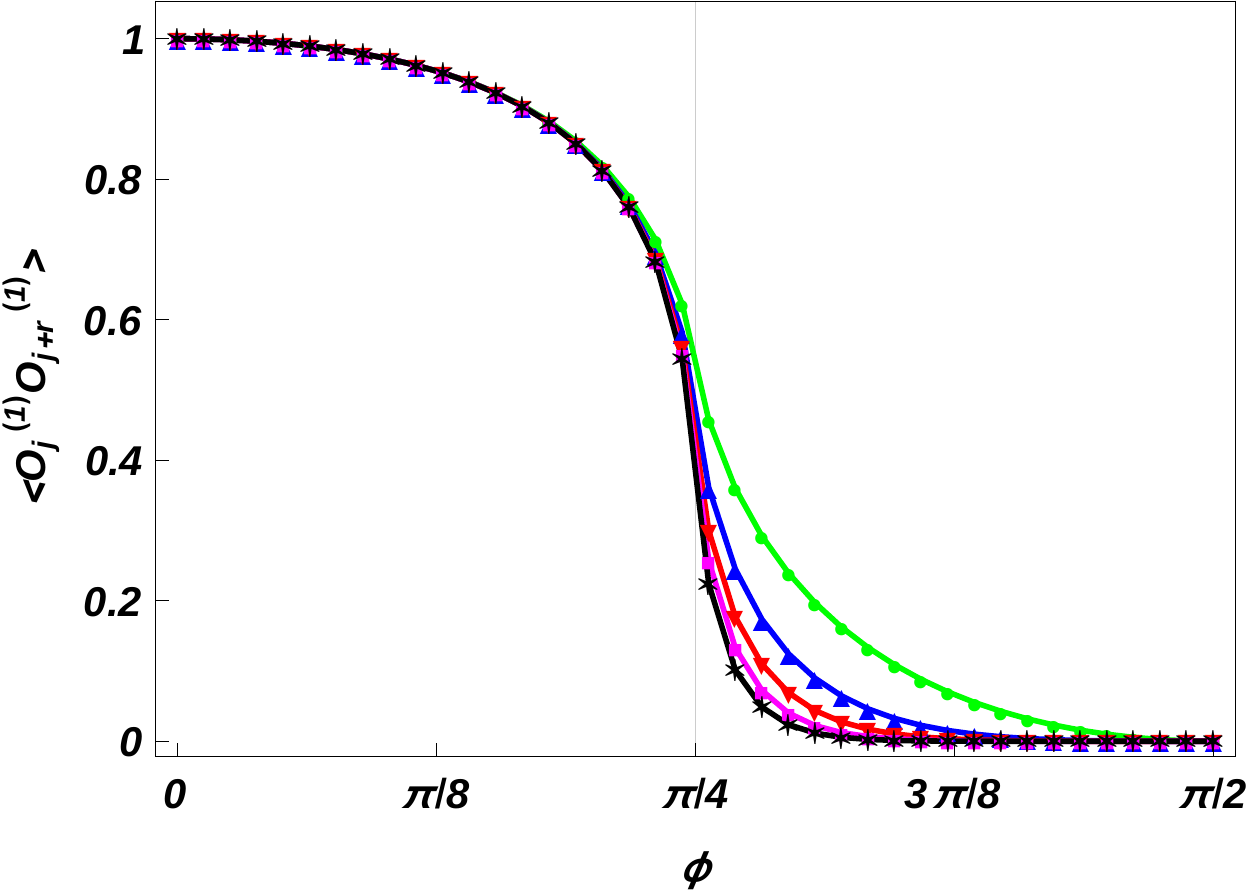}
\caption{(Color online) Behavior of the expectation value $\big \langle \mathcal{O}_j^{(n)} \mathcal{O}_{j+r}^{(n)}\big \rangle$, 
for $n=1$ and $r=3,6,9,12,15$, as a function 
of the phase parameter $\phi$: green dots (upper curve) $r=3$, blue up-triangles $r=6$, red down-triangles $r=9$, magenta squares $r=12$ 
and black stars (lower curve) $r=15$. 
As $r$ increases, the expectation value tends to disappear in the paramagnetic phase while it remains finite in the cluster phase.}
\label{stringorderparameterfig1}
\end{figure}

\begin{figure}[t]
\includegraphics[width=0.42\textwidth]{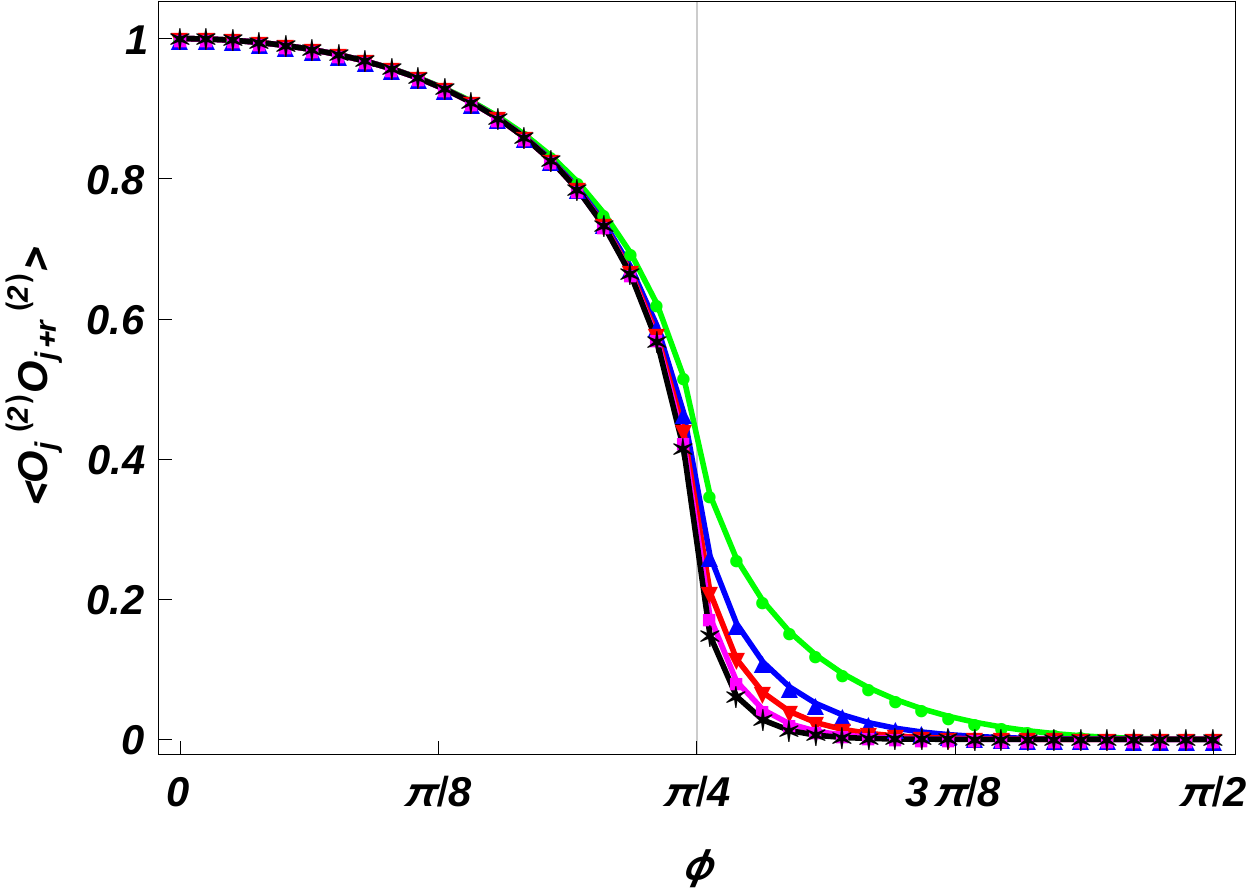}
\caption{(Color online) Behavior of the expectation value $\big \langle \mathcal{O}_j^{(n)} \mathcal{O}_{j+r}^{(n)}\big \rangle$, 
for $n=2$ and $r=3,6,9,12,15$, as a function 
of the phase parameter $\phi$: green dots (upper curve) $r=3$, blue up-triangles $r=6$, red down-triangles $r=9$, magenta squares 
$r=12$ and black stars (lower curve) $r=15$. 
As $r$ increases, the expectation value tends to disappear in the paramagnetic phase while it remains finite in the cluster phase.}
\label{blockmagnetizationfig1}
\end{figure}

\begin{figure}[t]
\includegraphics[width=0.42\textwidth]{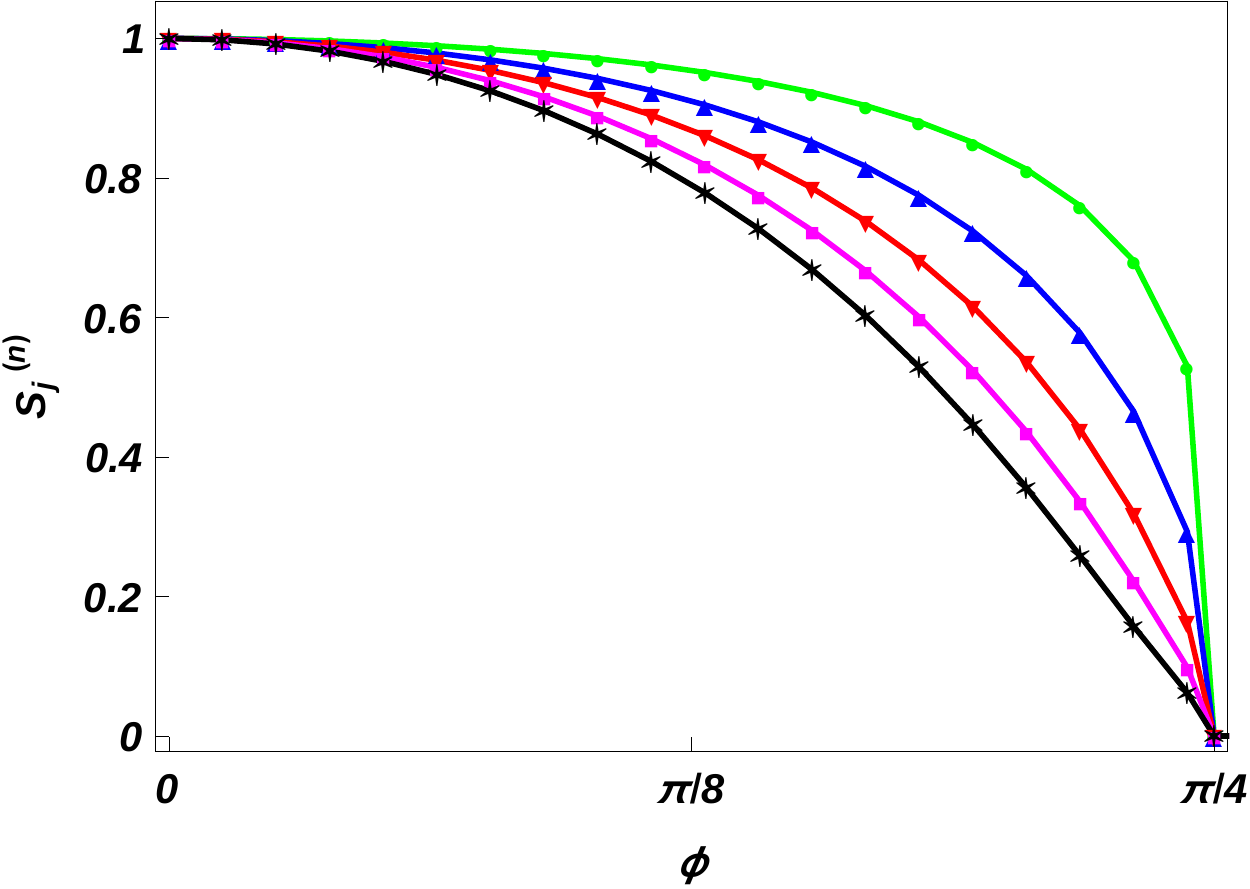}
\caption{(Color online) Behavior of the string order parameter $S_j^{(n)}$, for $n=1,3,5,7,9$, as a function of the phase parameter $\phi$: green dots (upper curve) $n=1$, 
blue up-triangles $n=3$, red down-triangles $n=5$, magenta squares $n=7$ and black stars (lower curve) $n=9$. The dots represent the numerical
results of the string order parameter $\mathcal{S}_j^{(n)}$ given in eq.~(\ref{order_parameters}), whereas the curves correspond to the behavior of the string 
order parameter $S^{(n)}$ defined in eq.~(\ref{BneOn}).}
\label{stringorderparameterfig}
\end{figure}

\begin{figure}[b]
\includegraphics[width=0.42\textwidth]{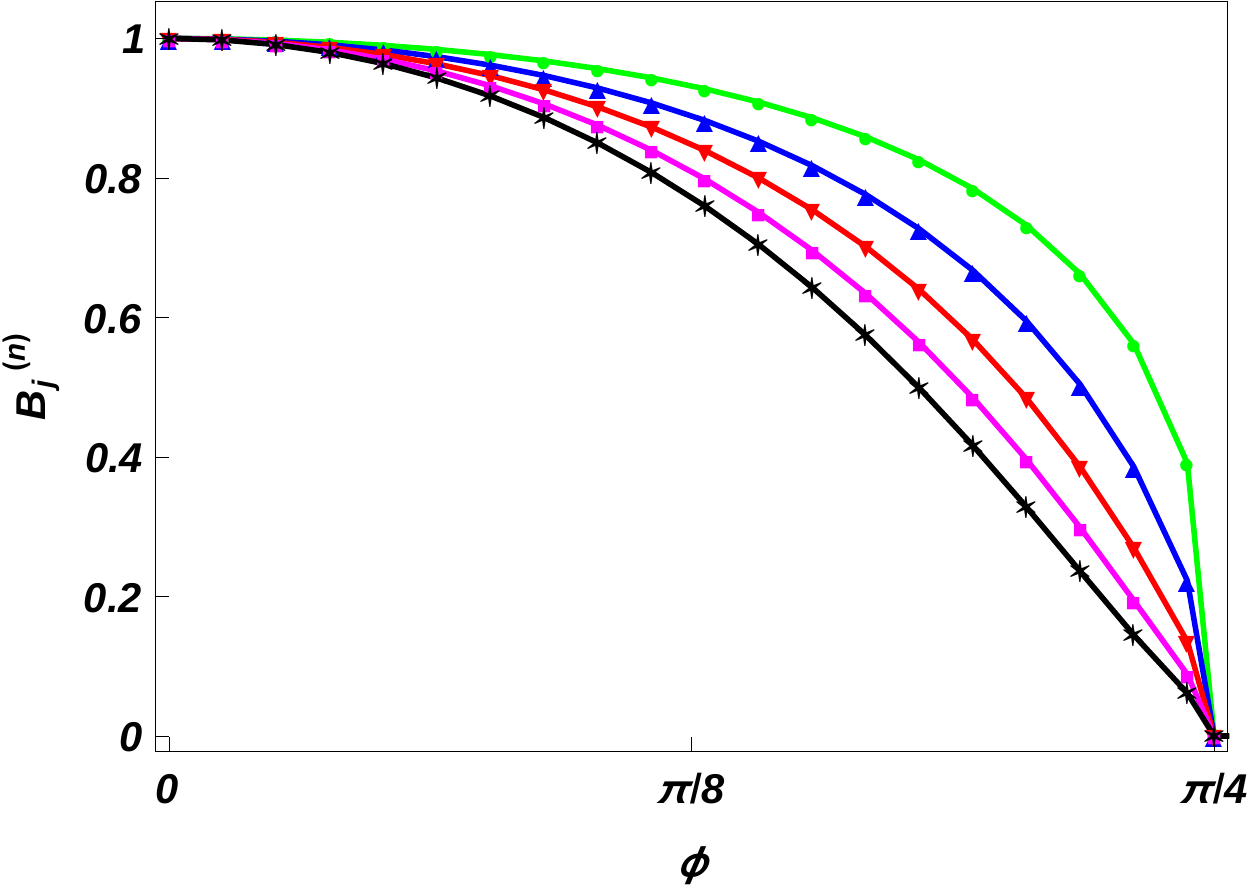}
\caption{(Color online) Behavior of the nematic order parameter $B_j^{(n)}$, for $n=2,4,6,8,10$, as a function of the phase parameter $\phi$: green dots (upper curve) $n=2$, 
blue up-triangles $n=4$, red down-triangles $n=6$, magenta squares $n=8$ and black stars (lower curve) $n=10$. The dots represent the
numerical results of the nematic order parameter $\mathcal{B}_j^{(n)}$ given in eq.~(\ref{order_parameters}), whereas the curves correspond to the behavior ot the nematic 
order parameter $B^{(n)}$ defined in eq.~(\ref{BneOn}).}
\label{blockmagnetizationfig}
\end{figure}

As we have seen in Sec.~\ref{Solution}, 
the behavior of the second derivative of the ground state energy density shows that, regardless the value of $n$, 
the system undergoes to a quantum phase transition at $\phi=\phi_c\equiv \pi/4$. 
However, the free energy does not detect the kind of phases realized. 
In this section, we determine the nature of the two phases, by studying the behavior of the order parameters that characterize them.
To make this analysis it is useful to rewrite the Hamiltonian in eq.~(\ref{Hamiltonian}) by using a well tailored cluster operators.
Such operators, that we named $\mathcal{O}_j^{(n)}$ are defined as 
\begin{eqnarray}
\label{varying_r}
\!\!\!\mathcal{O}_j^{(n)}\!\!\!&=&\!\!\! \left( \bigotimes_{k=1}^{j-n-1}\sigma_k^z\!\right) \sigma_{j-n}^y \sigma_{j-n+1}^x  
\cdots \sigma_{j-1}^y \sigma_{j}^x  \; \; \; \text{odd} \; n 
\nonumber \\
\!\!\! \mathcal{O}_j^{(n)}\!\!\!&=&\!\!\sigma_j^x \sigma_{j+1}^y \sigma_{j+2}^x \cdots \sigma_{j+n}^x \;\;\;\;\;\;\;\;\;\;\;\;\;\;\;\; 
\; \;\; \; \; \; \; \; \; 
 \text{even} \; n 
\end{eqnarray}
Hence, if $n$ is even, the operator $\mathcal{O}_j^{(n)}$ is defined on a finite support made by $n+1$ contiguous spins while for odd $n$ 
the dimension of support is infinite.
A fundamental properties of $\mathcal{O}_j^{(n)}$ is that, fixing $n$ $[\mathcal{O}_{j}^{(n)},\mathcal{O}_{k}^{(n)}]=0 \; \forall j,\; k$.
By using this operator we can rewrite the Hamiltonian of eq.~(\ref{Hamiltonian}) as 
\begin{equation}
\label{Hamiltonian4}
 H_{\phi}^{(n)}=-J \cos(\phi)\!\sum_{j}\! \mathcal{O}_{j}^{(n)} \mathcal{O}_{j+1}^{(n)}\!+\!J \sin(\phi)\!  \sum_{j}\! \sigma_{j}^z \; .
\end{equation}
Starting form this new expression of the Hamiltonian in eq.~(\ref{Hamiltonian}), and taking into account the commutation property of the operators
$\mathcal{O}_j^{(n)}$, it is easy to argue that, for $\phi<\phi_c$ the order parameter is given from the expectation value of $\mathcal{O}_j^{(n)}$. 
Hence, in agreement with this hypothesis we have that below $\phi_c$ our family of models shows, depending on $n$, two very different phases.
One, for even $n$ in which the order parameter is defined on a finite support made by $n+1$ spins (nematic phase), and the second, for odd $n$, 
that admits an order parameter with an infinite support (symmetry protected topological phase).
This hypothesis is strengthened if we look at Fig.~\ref{stringorderparameterfig1} ($n$ odd) and Fig.~\ref{blockmagnetizationfig1} ($n$ even) in which
we plot the expectation values of$\big \langle\mathcal{O}_{j}^{(n)} \mathcal{O}_{j+r}^{(n)}\big \rangle$ as a function of $\phi$, by varying $r$. 
From the two figures, we observe that, as $r$ increases, these expectation values tend to disappear for $\phi>\phi_c$, while they remain finite in 
the cluster phase, making $\mathcal{O}_{j}^{(n)}$ a perfect candidate for the order parameter. 
Accordingly with the usual approach~\cite{BMD1970,BM1971,GH2015} the order parameter, in the two cases, is defined as  
\begin{eqnarray}
\label{order_parameters}
\mathcal{S}_j^{(n)}&=&\sqrt{\lim_{r \to \infty} 
\Big \langle\mathcal{O}_{j}^{(n)} \mathcal{O}_{j+r}^{(n)}\Big \rangle} \; \; \; \;\;\;\;\text{odd} \;\; n \; , \nonumber \\
 \mathcal{B}_j^{(n)}&=&\sqrt{\lim_{r \to \infty}  
 \Big \langle\mathcal{O}_{j}^{(n)} \mathcal{O}_{j+r}^{(n)}\Big \rangle} \;\;\;\;\;\;\; \text{even} \;\; n \; .
\end{eqnarray}
In Fig.~\ref{stringorderparameterfig} and Fig.~\ref{blockmagnetizationfig} we show the numerical results for the order parameter in the two cases.
Therefore we can claim that for an even value of $n$, the system is in a nematic phase, characterized by an order parameter ($\mathcal{B}_j^{(n)}$) 
defined on a block of spins with dimension equal to $n+1$, while for an odd value of $n$, the system is in a symmetry-protected topological ordered 
phase, characterized by a string order parameter ($\mathcal{S}_j^{(n)}$).
Analyzing the numerical data obtained for both order parameters $\mathcal{S}_j^{(n)}$ and $\mathcal{B}_j^{(n)}$, we find finally the same dependence 
on $n$ and $\phi$, i.e.
\begin{eqnarray}\label{BneOn}
 \mathcal{S}^{(n)}&=&\left(1-\tan(\phi)^{2}\right)^{\frac{n+1}{8}} \; , \nonumber \\
 \mathcal{B}^{(n)}&=&\left(1-\tan(\phi)^{2}\right)^{\frac{n+1}{8}} \; .
\end{eqnarray}
from which we deduce the critical exponent $\beta$
\begin{equation}
 \label{beta}
 \beta=\beta(n)=\frac{n+1}{8}\;,
\end{equation}
that depends on $n$. 
This fact point out that our models, depending on $n$, fall into different classes of symmetries.
On the contrary, above the quantum critical point, i.e. $\phi>\phi_c$, the system is dominated by the external magnetic field. In such a phase, 
there is no order parameter and the system is in a typical paramagnetic phase.

\section{The entanglement properties}
\label{Entanglementproperties}

In this section, we analyze the entanglement properties among spins as well as between a block of spins and the rest of the chain. 
Despite the complexity of the class of models under investigation, we obtain general results, that show the relevance of the entanglement features 
in these complex quantum systems. 
From the analysis of the block entanglement we will be able to determine the central charge of the model~\cite{CC2004} while the analysis of the 
entanglement of the spins allows us to unveil the role of the interplay between cluster and Ising interaction to produce both bipartite and 
multipartite  entanglement.

As usual, the study of the entanglement properties starts from the analysis of the reduced density matrix of $m$ spins, obtained by tracing out all 
the degrees of freedom of the remaining spins of the system. 
The reduced density matrix decomposes in terms of the $m$-point spin correlation functions
\begin{equation}
 \label{densitymatrix}
 \rho_m^{(n)}=\frac{1}{2^m} \sum_{\alpha_1,\dots,\alpha_m} \langle \sigma_1^{\alpha_1}\sigma_2^{\alpha_2}\cdots\sigma_m^{\alpha_m}\rangle
 \sigma_1^{\alpha_1}\sigma_2^{\alpha_2}\cdots\sigma_m^{\alpha_m}\; ,
\end{equation}
where $\alpha_i=0,x,y,z$ and $\sigma_i^{0}$ denotes the identity matrix.

We consider three different measures of entanglement: 1) the entanglement between two spins quantified by the 
concurrence~\cite{Concurrence}; 2) the genuine multipartite entanglement~\cite{HH2008,HHBE2012,MCCSGH2011} between spins in a block; 
3) and the the entanglement between a block of adjacent spins and the rest of the chain, as quantified by the von Neumann entropy. 

\subsection{Entanglement between two spins}

For what concerns the entanglement between two spins in a block, we prove the following theorem: 

\begin{theorem}
If the distance $r$ between the two spins is not an integer multiple of $n+1$ the two spins are not entangled
\end{theorem}

\textbf{Proof:} To proof this theorem is it enough to recall the results obtained in Sec.~\ref{Spincorrelation}, for the spins correlation functions.
In fact, in agreement with eq.~(\ref{densitymatrix}), the $2$-spin reduced density matrix can be written as a linear composition of single-body and 
two-body spin correlation functions. 
For what concern the single-body correlation functions, we have that $\langle \sigma_i^x\rangle=\langle \sigma_i^y\rangle=0$, because of the 
properties of the Majorana fermionic operators $\langle A_i\rangle=\langle B_i\rangle=0$.
On the other hand, for what concern the two-body correlation functions, all the functions that involve different spin operators vanish in 
agreement with the fact that $\langle A_i\rangle=\langle B_i\rangle=0$ and $\langle A_i A_j\rangle=\langle B_i B_j\rangle=0$ if $i \neq j$. 
Thus, the two spin reduced density matrix depends on four different correlation functions only: $\langle \sigma_i^z\rangle$,  
$\langle \sigma_i^x \sigma_{i+r}^x\rangle$, $\langle \sigma_i^y \sigma_{i+r}^y\rangle$ and $\langle \sigma_i^z \sigma_{i+r}^z\rangle$. 
However, $\langle \sigma_i^x \sigma_{i+r}^x\rangle=\langle \sigma_i^y \sigma_{i+r}^y\rangle=0$, if $r \neq l(n+1)$ with $l$ integer. 
Hence, the reduced density matrix depends only on $\langle \sigma_i^z\rangle$ and $\langle \sigma_i^z \sigma_{i+r}^z\rangle$ and therefore is 
diagonal in the basis of the eigenstates of $\sigma_i^z$ and $\sigma_{i+r}^z$. 
Being diagonal in a base made by states that are tensor product of local states, the reduced density matrix cannot be entangled. Q.E.D.

\begin{figure}[t]
\includegraphics[width=0.43\textwidth]{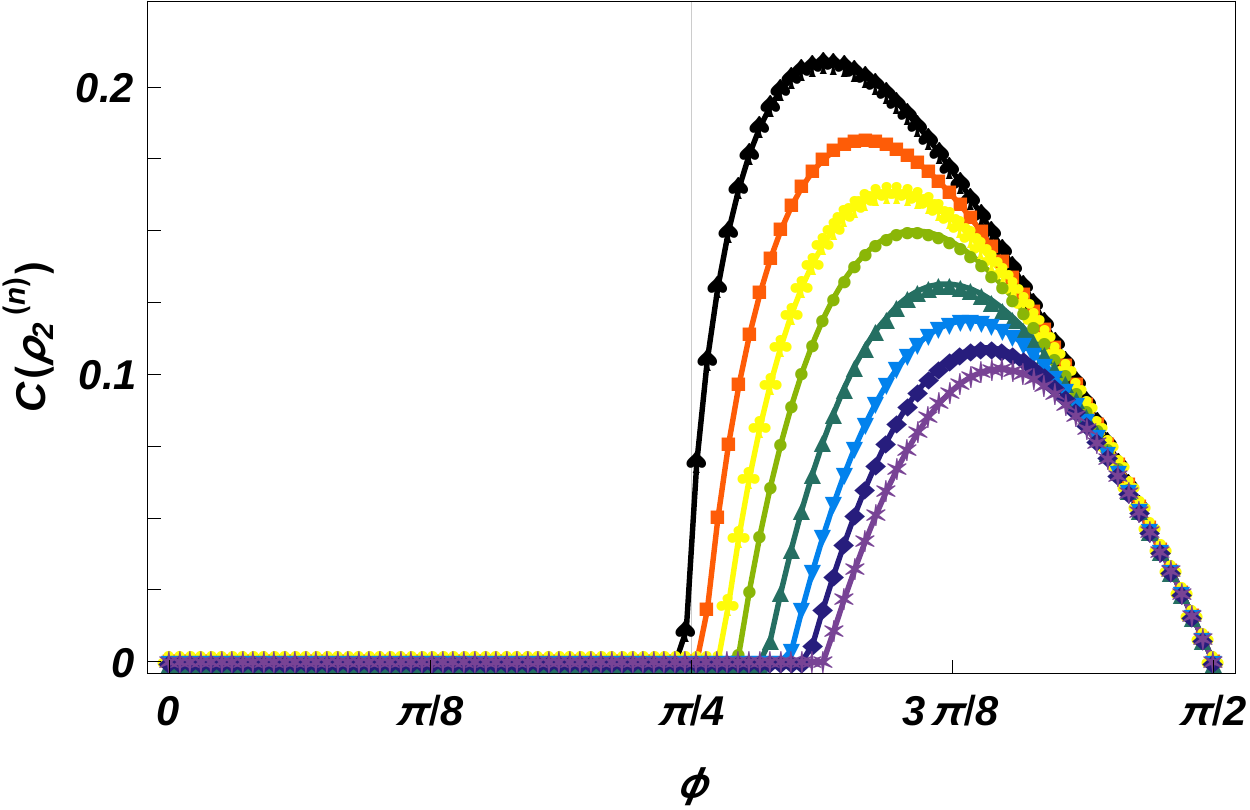}
\caption{(Color online) Dependence of the concurrence $C(\rho_2^{(n)})$ as function of the phase parameter $\phi$, for $r=n+1$ and different $n$ 
that runs from $1$ 
(highest curve) to $12$ (lowest curve). Note that only for $n=1$ concurrence is non-zero before and after the critical point and, generally, 
it decreases with increasing 
cluster size.}
\label{concurrence1}
\end{figure}

On the contrary, when $r=l(n+1)$, since $\langle \sigma_i^x \sigma_{i+r}^x\rangle\neq0$ and $\langle \sigma_i^y \sigma_{i+r}^y\rangle\neq0$, 
the reduced density matrix is not classical and it exists a region of the Hamiltonian parameters for which spins are entangled. 
We quantify such entanglement in terms of the concurrence $C(\rho_2^{(n)})$~\cite{Concurrence}. 
In Fig.~\ref{concurrence1}, we plot $C(\rho_2^{(n)})$ as a function of the phase parameter $\phi$, for $r=n+1$ and for different values of $n$.
For each $n$ the concurrence shows a similar behavior: it is different from zero in a region confined in the paramagnetic phase, with the only 
exception of $n=1$; increasing $n$, the concurrence becomes smaller and smaller and the relative maximum goes towards higher value of $\phi$. 
However, at  $\phi=\pi/2$, regardless the value of $n$, the systems admit a factorization point~\cite{GAI2008,GAI2009,GAI2010}. 
On the other hand, for all $l>1$ we have that all the concurrences are identically zero. Therefore, the entanglement is always limited between spins 
at the ends of the cluster.

\subsection{Genuine Multipartite entanglement}

For what concern the genuine multipartite entanglement, we prove the following theorem:

\begin{theorem}
 For each block made by $m$ adjacent spins, with $m\le n+2$, there is no genuine multipartite entanglement
\end{theorem}

\textbf{Proof:}
The proof is based on the fact that, following the definition of the genuine multipartite entanglement for a mixed state, it must be impossible to 
find a decomposition of the reduced density matrix in states that show only entanglement between a couple of spins. 

To start our proof, let us consider  a block made by $m=n+2$  adjacent spins
The reduced density matrix on such a block can be written in terms of the spin correlation functions and, in turn, 
all the spin correlation functions must be written in terms of the $G_r(n,\phi)$ functions. 
Taking into account the results of Sec.~\ref{Spincorrelation} and the fact that the 
maximum distance between the two spins in the block is $n+1$, the reduced density matrix depends only on three different fermionic correlation 
functions: $G_{0}(n,\phi)$,  $G_{n+1}(n,\phi)$ and $G_{-n-1}(n,\phi)$. 
Therefore, the only spin correlation functions different from zero are that diagonal in the natural basis or that associated to an 
inversion of the two spins at the end of the block. Hence, in the natural basis, the reduce density matrix can be written as a convex combination 
\begin{equation}
 \label{decomposition}
 \rho^{(n)}_{n+2}=\sum_{i} p_i \left(\bigotimes_{k=2,n+1} \chi_{i,k} \right)\bigotimes  \chi_{i,1,n+2} \, ,
\end{equation}
where $\chi_{i,k}$ is a state defined on the $k$-th spin of the block and $\chi_{i,1,n+2}$ is a state (entangled or not) defined on the two 
endpoints of the block. 
In other words, the reduced density matrix can be written as a sum of states that, with the only exception of a possible bipartite entanglement 
between the two endpoints of the block, are fully factorized.
In such state it comes immediately that any multipartite entanglement vanishes. 

If now we consider a block made by $m=n+1$ adjacent spins, we have that the reduced density matrix can be obtain by eq.~(\ref{decomposition}) 
tracing out one of the two endpoint spins. Thus, we obtain a reduced density matrix that is a linear convex combination
of fully disentangled states and hence it does not admit any entanglement. 
Moreover, also any subsystem made by $m<n+1$ spins cannot show multipartite entanglement. Q.E.D.  

It is interesting to make a comparison with the results reported in Ref.~\cite{GH2015}, for the $n$-cluster Ising models, where, on the contrary, 
there is no bipartite entanglement but a significant value of genuine multipartite entanglement, confined in the anti-ferromagnetic phase, 
with the exception of $n=1$. 
Comparing these two results, and taking into account the proof of the presence of genuine multipartite entanglement in the 
$xy$-model~\cite{GH2013,HOG2014}, we may counter-intuitively conclude that a fundamental requirement to have multipartite entanglement
is the presence, in the Hamiltonian, of a simple Ising-like interaction.  

\subsection{Block entanglement}

Another important property in multipartite systems concerns the entanglement between a block of $m$ adjacent spins and the rest of the chain, 
and its relation to the holomorphic and anti-holomorphic sectors in conformal field theory~\cite{CC2004}.

For this purpose, we compute the von Neumann entanglement entropy of the reduced density matrix of a block of $m$ spins
\begin{equation}
\label{VNEdefinition}
 S(\rho_{m}^{(n)})=-\text{Tr}\left[\rho_m^{(n)} \log_2(\rho_m^{(n)})\right]\;.
\end{equation}
Using the methods developed in Ref.~\cite{VLRK2003,LRV2004}, we find
\begin{equation}
 \label{VNEdefinition1}
 S(\rho_{m}^{(n)}) = \sum_{j=1}^m H_{\textrm{Shannon}}\left(\frac{1+\nu_j^{(n)}}{2}\right) \; ,
\end{equation}
where $H_{\textrm{Shannon}}(x)$ is the Shannon entropy
\begin{equation}
 \label{Shannonentropy}
 H_{\textrm{Shannon}}(x)= -x \log_2(x) -(1-x) \log_2(1-x)\; ,
\end{equation}
and $\nu_j^{(n)}$ are the imaginary part of the eigenvalues of the matrix
\begin{equation}
\label{gammaprime}
 \left(\Gamma^{(n)'}\right)_{ij}= \delta_{ij}-\imath \left(\Gamma_m^{(n)}\right)_{ij} \; ,
\end{equation}
with
\begin{equation}
\label{gamma}
 \Gamma_m^{(n)}=\left( \begin{array}{cccc}
 \vspace{0.1cm}
 \! \Pi_0^{(n)} & \Pi_{-1}^{(n)} &\! \cdots \!&  \Pi_{-m+1}^{(n)} \! \\
 \! \Pi_{1}^{(n)} & \Pi_0^{(n)} & \!\cdots \!&  \Pi_{-m+2}^{(n)}\! \\
 \! \vdots & \vdots & \!\ddots \!& \vdots\! \\
 \! \Pi_{m-1}^{(n)} &\Pi_{m-2}^{(n)} & \!\cdots\! & \Pi_0\!
 \end{array}
\right) \; ,
\end{equation}
and
\begin{equation}
\label{pi}
 \Pi_r^{(n)}=\left(
 \begin{array}{cc}
 0 & G_{r}(n,\phi) \\
 -G_{-r}(n,\phi) & 0 \\
 \end{array}
\right)\;.
\end{equation}

\begin{figure}[t]
\includegraphics[width=0.44\textwidth]{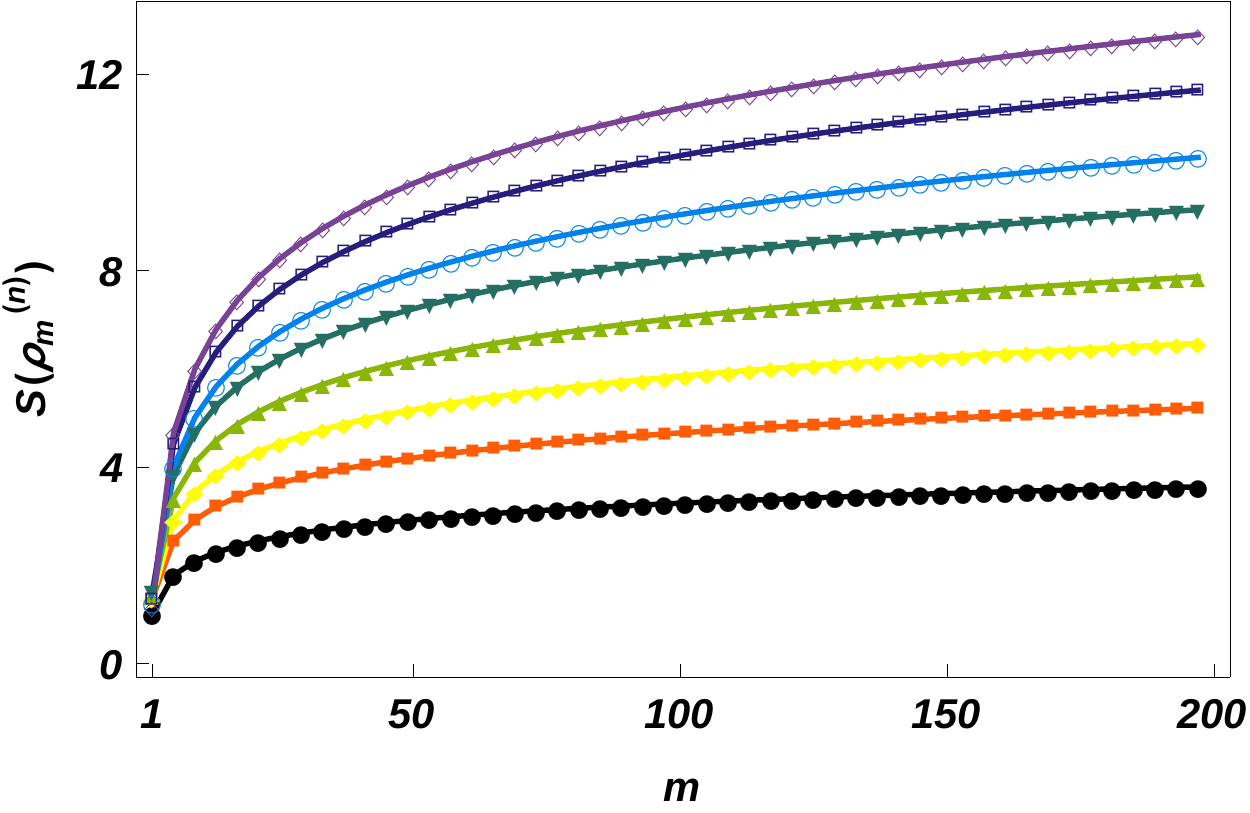}
\caption{(Color online) Behavior of the von Neumann entropy $S(\rho_{m}^{(n)})$, as a function of the size $m$ of the block, for different values 
of size $n+2$ of the 
cluster. The value of $n$ runs from $1$ (lowest black curve) to $8$ (highest violet curve). Independently of $n$, the von Neumann entanglement 
entropy diverges, at a quantum critical point $\phi_c=\pi/4$, as a logarithmic function of $m$.}
\label{voneumannfig}
\end{figure}

We evaluate numerically the von Neumann entanglement entropy as a function of the size $m$ of the block, at the critical point $\phi=\phi_c$, for
$n$ that runs from $1$ to $8$ and plot the results in Fig.~\ref{voneumannfig}.

Analyzing the numerical data, we deduce
\begin{equation}
\label{numericalfitvonneumann}
 S(\rho_{m}^{(n)})\simeq0.17(1+n)\; \log_2 m+\text{const}(n) \; .
\end{equation}
The multiplicative constant in front of the logarithmic term is known to be related to the central charge of the $1+1$ dimensional conformal field 
theory, that describes the critical behavior of the chain via the relation~\cite{HLW1994}
\begin{equation}
 \label{cft}
 S_m = \frac{\mathsf{c}+\overline{\mathsf{c}}}{6}\; \log_2 m\;,
\end{equation}
where $\mathsf{c}$ and $\overline{\mathsf{c}}$ are the central charges of the so-called holomorphic and anti-holomorphic sectors. 
Due to the existence of a duality in the system, we have that $\mathsf{c}=\overline{\mathsf{c}}$ and hence
\begin{equation}
\label{centralcharge}
  \mathsf{c}=\mathsf{c}(n)\simeq 0.51 (1+n)\;.
\end{equation}
Two quantum one-dimensional systems belong to the same universality class if they have the same central charge. 
In this case, the central charge depends on $n$. This implies that the $n$-cluster models in a transverse magnetic field fall into 
different classes with respect to their symmetries.

\section{Conclusions}
\label{Conclusions}

In summary, we analyzed a family of fully analytically solvable models, named $n$-cluster models in a transverse magnetic field. 
These  models are characterized by a $n+2$-body cluster interaction competing with a spatially uniform transverse magnetic field. 
Using the Jordan-Wigner transformations, we diagonalized the models and proved that their classes of symmetry depends on $n$. 
However, regardless the value of $n$, a phase transition always occurs exactly when both interactions are equally weighted. 
The paramagnetic phase, realized for $\phi>\phi_c$, shows a very similar behavior for all $n$. 
On the contrary, the cluster phase, realized for $\phi<\phi_c$, exhibits two different orders, depending on $n$. 
For odd or even cluster size $n+2$, we have a symmetry-protected topological ordered phase or a nematic ordered one respectively, in agreement 
with the results obtained in Ref.~\cite{GH2015}. 

We also investigated how the complexity of the orders translates to the entanglement properties. 
In completely contrast with the results obtained for the $n$-cluster Ising models~\cite{GH2015}, any possible multipartite entanglement vanishes, 
while the bipartite entanglement, quantified in terms of the concurrence between two spins at a distance $n+1$, has a non vanishing value in a region 
confined in the paramagnetic phase, with the only exception of $n=1$.

The importance of this family of fully analytically solvable models is the presence of exotic phases, such as nematic and topological phases. 
Hence, it can provide a new field to study hot topics of the current research, as the presence of global entanglement~\cite{HGI2016} or the 
effects of a sudden quench of the Hamiltonian parameters, for ground states that violate the symmetries of the system~\cite{GZ2016}.

Moreover, this family of models can be generalized with respect to higher dimensions, both in space and degrees of freedom, and may become a
prototype for studying the 
possible applications of quantum information tasks.

\section*{Acknowledgments} 

We wish to thank M. Dalmonte for the interesting discussions. G. Z. thanks 
EQuaM - Emulators of Quantum Frustrated Magnetism, Grant Agreement No. 323714  while
S.M.G. acknowledges financial support from the Ministry of Science, Technology and Innovation of Brazil

\end{document}